\documentclass{article}

\usepackage{graphicx}
\usepackage{longtable}
\usepackage{amsmath}
\usepackage{amssymb}
\usepackage{amsfonts}
\usepackage{framed}
\usepackage{fancyvrb}

\def\lognp{log $N$-log $P$}
\def\logns{log $N$-log $S$}
\def\cmaxcmin{$C_{\rm max}/C_{\rm min}$}
\def\cmin{$C_{\rm min}$}
\def\cmax{$C_{\rm max}$}

\title{THE PIONEER VENUS ORBITER CATALOG OF GAMMA-RAY BURSTS}
\author{E.~E.~Fenimore$^1$, A.~Crider$^2$, J.~J.~M.~in'T Zand$^3$,\\
 R.~W.~Klebesadel$^1$, J.~G.~Laros$^1$, and M.~Meier$^1$}
\date{%
    $^1$Los Alamos National Laboratory\\%
    $^2$Elon University\\%
    $^3$Netherlands Institute for Space Research\\[2ex]%
     LA-UR 23-21872
}

\begin{document}

\maketitle

\begin{abstract}
The Pioneer Venus Orbiter (PVO) Gamma-ray burst experiment detected 318
gamma-ray bursts over about 14 years between 1978 and 1992 with near $4\pi$ coverage.  
This data set complements BATSE
by determining the properties of the brightest GRBs.  PVO places a
constrains on the slope of the bright end of the \lognp~distribution.
The slope is -1.52$\pm 0.15$. 
\end{abstract}

\section{INTRODUCTION}

The discovery of gamma-ray bursts (GRBs, Klebesadel et al.~1973) led
to decades of intense study to determine
their nature.  
The Burst And Transient Source Experiment (BATSE)
showed that the distribution of GRBs is
 isotropic yet inhomogeneous implying that
they are further away than typical galactic distances.
The discovery of optical transients associated with the bursts (Costa et al, 1997)
settled that the bursts came from far away galaxies.

The \lognp~distribution is a statistical vehicle for estimating
the homogeneity of the spatial distribution of GRBS.
Here, $P$ is some measure of the event's
brightness (e.g., the peak photons cm$^{-2}$ s$^{-1}$) and $N$ is
the integral number of observed events brighter than $P$.
It is assumed that $P$ is a standard candle such that the brightness of events further
away  scale as $R^{-2}$ where $R$ is the distance to the events.
Since the number of events assuming a homogeneous distribution scales as $R^3$, the \lognp~distribution is
expected to be a -3/2 power law ($N \propto P^{-3/2}$).
For events that are very far away, the expansion of the universe shifts
the spectrum out of the instrument's bandpass and affects the apparent
density of events resulting in a \lognp~distribution which deviates
from the -3/2 power law even for a homogeneous distribution.  
 The degree of deviation
can be used to estimate the distance to the events.
Unfortunately, there is no reason to believe any particular measure of $P$
is, indeed, a standard candle.  It is likely that source evolution and/or
an intrinsic luminosity function also affects the distance determination.

In this paper, we present a catalog of GRBs observed by the Pioneer
Venus Orbiter (PVO) gamma-ray burst detector.
The emphasis is on providing information that can be combined with other
data to produce a combined \lognp~distribution with a large
dynamic range.
The large exposure factor for PVO ($\sim 145$ yr steradian) but small size
(22 cm$^2$) complements BATSE's small exposure factor (16.8 yr ster) and
large size (2000 cm$^2$).  BATSE detects the most distant, faint but
numerous events whereas PVO observes the rare bright events.  It
would take a BATSE-like experiment 30 years to detect as many bright events as contained
in this catalog.
PVO establishes a \lognp~distribution consistent with a -3/2 power law. 

\section{INSTRUMENTATION}

The PVO instrumentation (Klebesadel et al.~1980) consists of two CsI
detectors each 3.81 cm in diameter and 3.175 cm in length.  A phoswich
arrangement
is used to reduce the background counting rate from particles.
The primary shielding is a layer of 0.025 cm thick lead.
The two detectors were mounted at opposite edges of the PVO satellite
and the signals were added together.
Usually, there is no knowledge of the angle of incidence of an event.
  However, due to the symmetry of the detectors,
the effective area is very similar in all directions.
PVO has four energy channels; 100-200, 200-500, 500-1000, and 1000-2000 keV.
PVO has two basic types of data: ``real time'' (RT) data and
``memory readout'' (MRO) data.
The RT data are available continuously whenever the satellite is being
tracked ($\sim$ 57\% of the time between Sept.~1978 and Jan.~1992).
The temporal resolution of the RT data varies depending on the
telemetry rate.  The RT data contain high temporal resolution,
full-energy-range samples (100-2000 keV) and the four energy-resolved channels
at a lower temporal rate.
There are four full-energy-range samples for each set of energy-resolved channels.
For example, during $\sim$ 12\% of the time, PVO had a
resolution of 0.125 s in the full-energy-range
and 0.5 s for the four energy channels.
Nearer the end of the mission, more time was spent with longer accumulations (e.g.,
8 s for the full energy range).

Statistically significant variations in the counting rate
cause data to be stored in  a special
memory with higher time resolution.
This memory is usually read out once per day and is referred to as the
Memory Readout (MRO) or ``triggered'' data.
The default temporal resolution of the MRO data is 11.72 ms for the
full energy range and 187.5 ms for the four-channel-spectral data.
Roughly 240 full energy range samples ($\sim 2.8$ s) before the
trigger and $\sim 2160$ samples (25.3 s) after 
the trigger are stored in the MRO.

\section{Catalog of Triggered PVO GRBs}

Tables 1 and 2 summarize the characteristics of 318 PVO triggered events.
These results are based on a complete reanalysis of the each event so the
values will differ from analyses prior to 1993.  In particular, we have
combined data from the RT and
MRO data to obtain the best possible information.
The following is a description of the columns in the tables.

The first two columns of Table 1 give the date and time of the event.
When multiple events occur on the same day, the brightest is designated
with an ``A'', the second brightest with a ``B'', etc.
The time is the universal time of the trigger at Venus. 
The third column ($\Delta T_{\rm RT}$) is the temporal resolution of the RT full energy range
channel in seconds.  If this column is blank, then only MRO data are available.

 The fourth column ($T_{\rm MRO}$) is the duration of the MRO record in seconds. The duration can be shorter
than 28.1 sec because either time-to-spill shorten the record (for bright bursts),
or the count rate satisfied the trigger-off criterion (for short
bursts), or only some of the 10 MRO blocks were available (denoted
by footnote I).

The fifth column (Bkg) presents the mean background counting rate in the
full-energy-range channel in counts/sec.
The next three columns give \cmaxcmin~for the three trigger
periods (1/4, 1, and 4 s).
Values of \cmaxcmin~greater than unity indicate that PVO would trigger on that
time scale.
When RT data are available, the telemetry includes a history of \cmin.
The telemetry does not include values of \cmax~or identify which
time period caused the trigger.
These were estimated in the ground processing.
For \cmax,
we used the peak count rate over $\Delta T_i$ samples
on the full energy range channel.
Each \cmax~was calculated once per $\Delta T_i$.  This introduces an
uncertainty in the true \cmaxcmin~since a different phase for $\Delta T_i$
could give a slightly different \cmax.
For \cmin, we used $11.2(B\Delta T_i)^{1/2}$ where Bkg is the third column of
Table 1.
The 11.2 is a trigger criterion parameter, effectively the trigger happens when
a sample is 11.2 standard deviations different than the background. 

In some cases, all three \cmaxcmin~were less than unity indicating that
the event should not have triggered PVO.  This sometimes happens because
of statistical variations or
the unknown phasing of the \cmax~samples.  There are a few cases (denoted
by footnote F) where the threshold trigger criterion was commanded to be 5.6 rather than 11.2.
The \cmaxcmin~were calculated with 11.2 standard deviations in all cases to allow a uniform
selection criterion.

The next two columns indicate the duration of the event in seconds.
The $T_{90}$ and $T_{50}$ values are the time that contains 90\% and
50\% of the total counts in the full-energy-range channel.

Table 2 provides information on the intensities of the events in two
energy bandpasses.
The 100 to 2000 keV bandpass is best suited for studies that use only PVO
data. Columns 2 and 3 use the energy range of PVO and are denoted with a subscript ``P'' for PVO.
Columns 4 to 7 scale the fluxes to the energy range of BATSE (50 to 300 keV). 
Columns 4 and 5 use the Band spectral shape (Band et al, 1996) to estimate the
fluxes in the 50 to 300 keV range and are denoted by the subscript "B" for Band.
Columns 6 and 7 use the thermal bremsstrahlung spectrum and are denoted by the subscript "T".

We provide peak values ($P$, in photons cm$^{-2}$ s$^{-1}$ above background, columns 2, 4, 6) 
evaluated for a sample interval of 250 ms and fluence
($S$, in $10^{-6}$ erg s$^{-1}$, columns 3, 5, 7)  values for the whole event.
To determine the peak values, we first locate the peak in the full energy
range channel with a sliding window of width 250 ms.  We then prorate the nearest
187.5 ms spectral samples to obtain the count spectrum corresponding to the
peak.
Various spectral shapes are fit (with a $\chi^2$ statistic) to the count
spectrum.
The resulting best fit spectral shape is integrated over either 100 to 2000
keV or 50 to 300 keV to obtain $P$.
The fluence is found in a similar manner except the count spectrum is
the net counts above background.
We have used two  spectral shapes: classic thermal
bremsstrahlung,
\begin{equation}
\phi(E) = E^{-1}\exp(E/E_0)~~,
\label{eqn:thermb}
\end{equation}
and the Band et al.~(1993) shape:
\begin{equation}
\phi(E) = \phi_0 E^\alpha \exp(-E/E_0)~~~~~~~~~~~~~{\rm if} (\alpha-\beta)E_0 > E \nonumber
\end{equation}
\begin{equation}
~~~~~~ = \phi_0 \big[(\alpha-\beta)E_0\big]^{\alpha-\beta} \exp(\beta-\alpha) E^{\beta}~~~~~~~~~{\rm if} (\alpha-\beta)E_0 < E~,
\label{eqn:band}
\end{equation}
Both  shapes give effectively the same PVO $P$ and $S$ values.
In table 2, columns 2 and 3
 give the $P_{\rm P}$ and $S_{\rm P}$ values for the 100 to 2000
keV bandpass based on the Band et al.~(1993) shape. Since PVO only has
four spectral channels, we fixed $E_0$ at 350 keV.  Thus, there were
three free parameters ($\phi_0, ~\alpha, ~\beta$).

To obtain values suitable for comparisons  to BATSE (50 to 300 keV),
 one must extrapolate
the PVO spectrum.
GRB spectra tend to change slope near 100 keV (see Band et al.~1993) so it
is difficult to accurately obtain values for 50 to 300 keV from PVO.
Equation \ref{eqn:band}
would
do the same thing if $E_0$ is a free parameter.  Thus, we fix $E_0$
to 350 keV.  The resulting peak intensity ($P_{\rm B}$) is in column 4 of Table 2.
Here, the subscript B indicates that the Band et al.~(1993) spectral shape was used to evaluate the intensity.
In column 6, we report $P_{\rm T}$, the peak intensity in 50 to 300 keV found
using the thermal bremsstrahlung shape (eq. \ref{eqn:thermb}).
The $P_B$ value probably gives a low estimate because the Band et al.~(1993)
 can
roll over more quickly than thermal bremsstrahlung. The thermal
bremsstrahlung shape probably gives a high estimate because that shape
cannot be flatter than $E^{-1}$ at low energy and sometimes GRB's spectra
are flatter.  One could use the average of
$P_{\rm B}$ and $P_{\rm T}$.  Footnote D indicates when, in our judgment, the
event burst data or background information was not complete
enough to produce reliable intensities.

The next two columns give a hardness ratio and the associated statistical
error.
The hardness ratio HR$_{\rm P}$ is based on
 the peak in the full-energy-range channel found with
a sliding 250 ms sampling window.
The error on HR$_{\rm P}$ (i.e., $\sigma_{\rm HR}$) is based solely on
Poisson statistics.
  The hardness ratio is defined to be the
counts above 200 keV to the counts below 200 keV.  The next column (HR$_S$)
is the hardness ratio for the whole burst.  No statistical error is provided
because the uncertainty is dominated by systematic effects such as the
background determination or the length of the available data.
These hardness ratios are provided only to give a rough
indication of the spectral nature of the
event.

\section{THE LOG ${\bf N}$-LOG ${\bf P}$/LOG ${\bf S}$ DISTRIBUTIONS}

\begin{figure}
\begin{center}
\includegraphics[width=\textwidth]{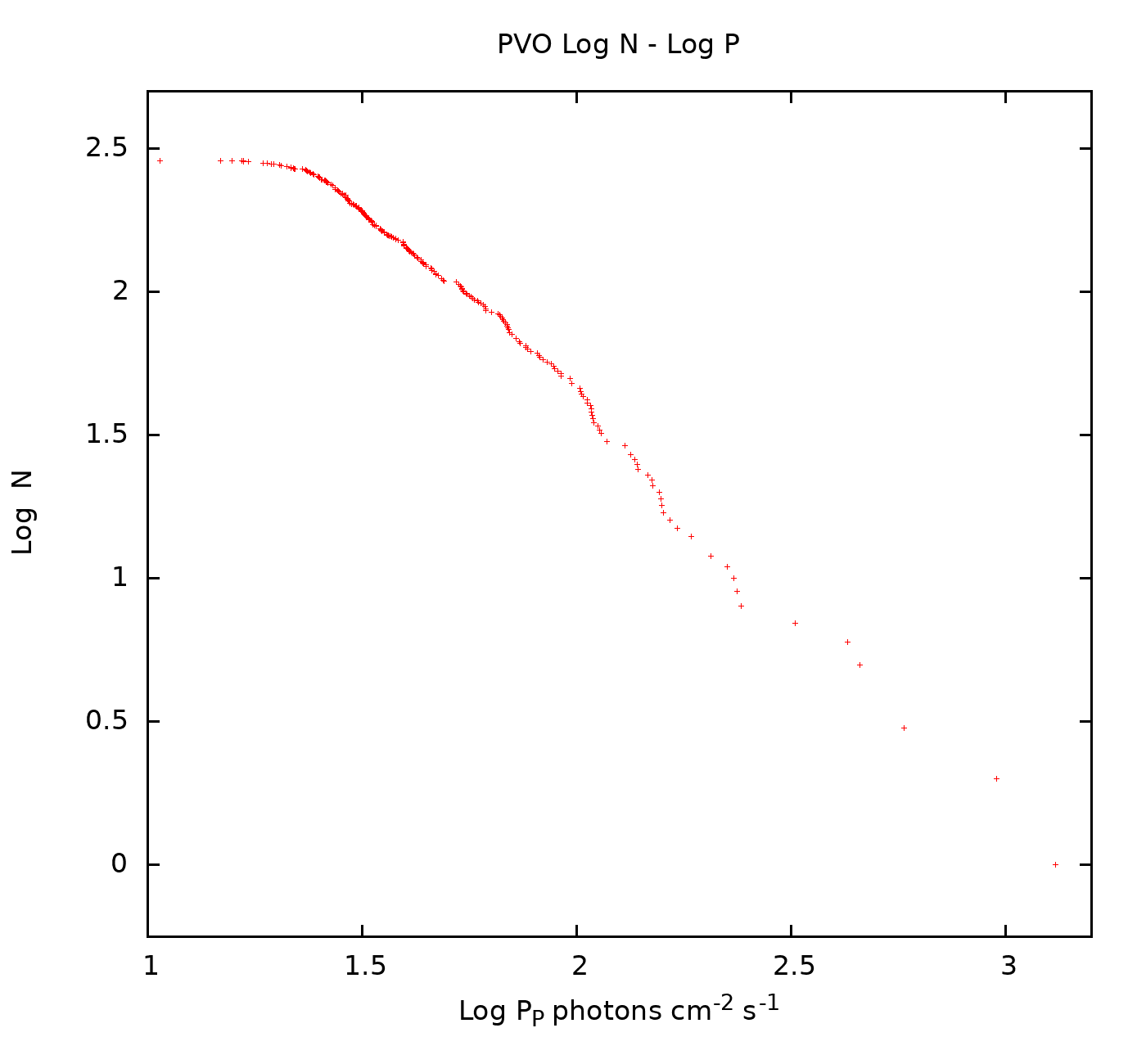}
\caption{The PVO Log N - Log P distribution. The peak luminosities are taken from the second column of Table 2, which
is the peak found over 250 ms in the PVO energy range of 100 to 2000 keV. The Band spectral shape and the PVO detector
response function were used to
integrate the spectrum over the energy range.  The 790305 event was omitted as it is not a GRB and as were events
with footnote "D" in Table 1.
} 
\label{fig:lognlogp}
\end{center}
\end{figure}

Figure \ref{fig:lognlogp}
shows the PVO \lognp~distribution based
on column 2 of
Table 2. To produce a uniform distribution, we omitted the March 5th event and events with footnote D.
The key issue regarding the PVO \lognp~is how well it constrains the
homogeneity of the local distribution.  We compare the observed \lognp~
distribution
to a power law ($N \propto P^{-\gamma}$) with a $\chi^2$ test.
We only used events with $P > 25$ photons cm$^{-2}$ s$^{-1}$ where the trigger efficiency
is effectively unity.
The  best fit slope is -1.52.
We estimate a confidence region on the slope with a Monte Carlo technique.
We produce 5000 randomly generated \lognp~distributions following
$N\propto P^{-\gamma}$ and fit them to a -1.52 power law.  We determine
what fraction (as a function of $\gamma$) of the randomly generated
data had a $\chi^2$ value as high as that observed with the actual data.
The  68\% confidence region is where the curves falls to $e^{-1/2}$
which occurs at
$1.52 \pm 0.15$.

\begin{figure}
\begin{center}
\includegraphics[width=\textwidth]{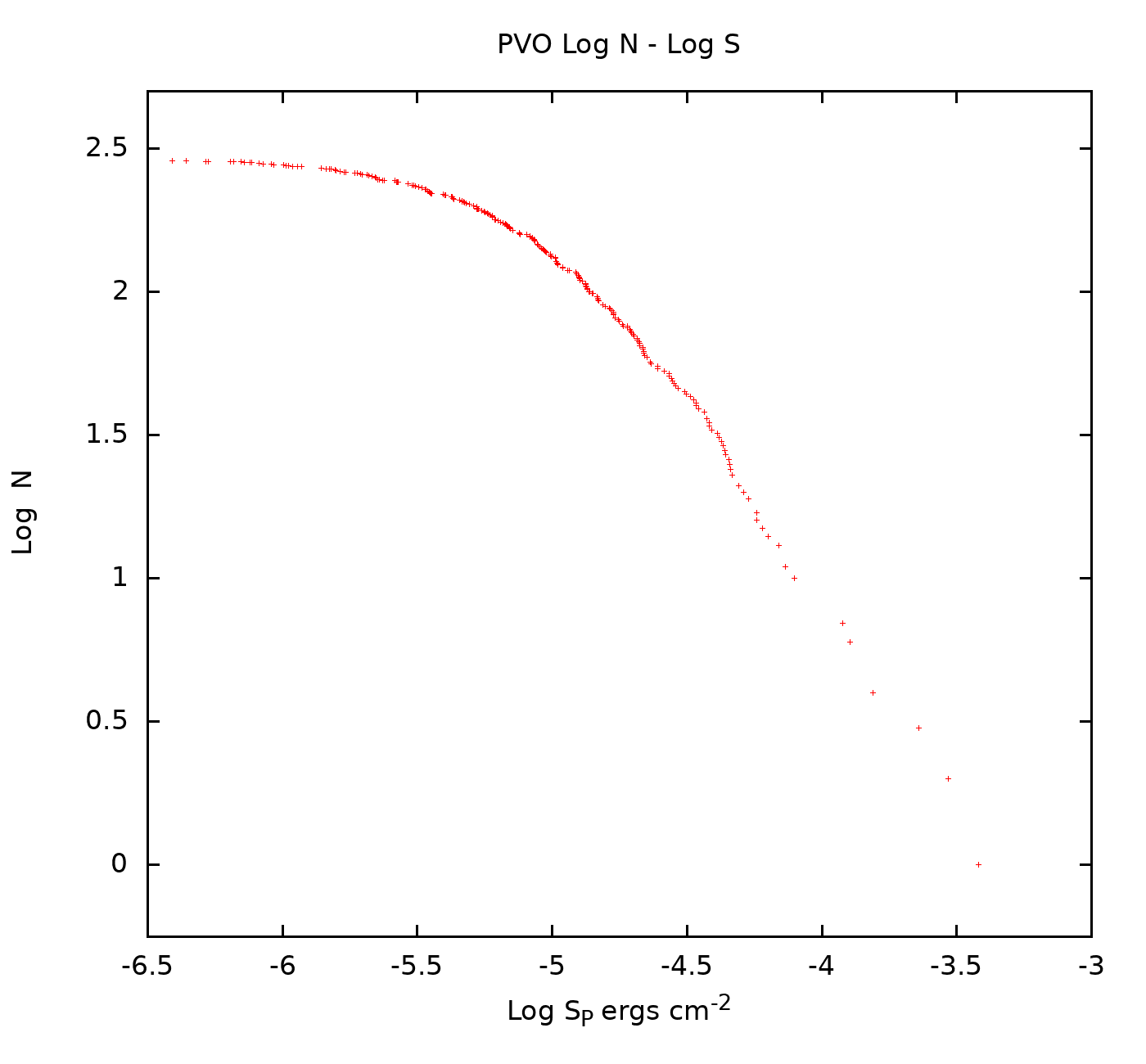}
\caption{The PVO Log N - Log S distribution. The fluences are taken from the third column of Table 2, which
is fluence in the PVO energy range of 100 to 2000 keV over the duration of the event. The Band spectral shape and the PVO detector
response function were used to
integrate the spectrum over the energy range.  The 790305 event was omitted as it is not a GRB as were events with footnote "D"
in Table 1.
} 
\label{fig:lognlogs}
\end{center}
\end{figure}

Figure \ref{fig:lognlogs} gives the \logns~distribution.  The \lognp~follows the -3/2 power law 
for a wider range of observations because the instrument triggers on $P$.

\section{SUMMARY}

In this catalog we have redetermined all parameters using the most complete
data that were available.  We have found the peak intensity (over 250 ms) and total fluence three different
way.  The $P_{\rm P}$ and $S_{\rm P}$ in Table 2 are found by fitting the energy-resolved PVO data to the Band shape and
integrating over
the PVO energy range of 100 to 2000 keV.
The $P_{\rm B}$ and $S_{\rm B}$ in Table 2 are found the same way except integrated over the BATSE energy range 
of 50 to 300 keV.
The $P_{\rm T}$ and $S_{\rm T}$ in Table 2 use the thermal bremsstrahlung spectrum (instead of 
the Band spectrum) and the BATSE energy range.
It is likely the the Band spectrum gives a low estimate and the thermal bremsstrahlung spectrum
gives a high estimate so an average might be more accurate.

To combine two experiments such as PVO and BATSE, one must establish a common selection criterion and
determine the exposure factors.
Events are selected within the instruments based on the trigger criteria.
The advantage of BATSE and PVO is that they both had
very stable trigger criteria.
BATSE triggers on timescales of 64, 256, and 1024 ms in the 50 to 300 keV
range.  PVO triggers on 250, 1000, and 4000 ms timescales in the 100 to
2000 keV range. 
Because both instruments trigger on (effectively) 1/4 and 1 s timescales,
one would accept only events that have \cmaxcmin~greater than unity on either
of those timescales.
In the case of PVO, one should exclude the March 5th 1979 event, events with footnote D,  and events in
1992.
In 1992, PVO was not tracked regularly and it is difficult to estimate
the correct exposure factor for that year.

Fenimore et al (1993) used the PVO data from this catalog combined with BATSE
with emphasis on the uniform selection of events. The uniform selection criteria,
when applied to both PVO and BATSE, cut the number of useful events in about half. 
 When uniformly selected, the BATSE data agrees with PVO to within the statistics. 
The combined \lognp~
distribution implies a standard candle luminosity of $6^{+1.3}_{-0.8} \times 10^{50}$ erg s$^{-1}$.  

This paper was drafted in 1993.

%
%

\pagebreak
\centerline{Table 1: Properties of PVO GRBs}
\begin{longtable}{l r r r r r r r r r r l}
\hline
\hline
Date &UT &$\Delta T_{\rm RT}$ &$T_{\rm MRO}$ &Bkg &1/4 s &1 s &4 s &TE &$T_{\rm 90}$ &$T_{\rm 50}$ &Notes \\
\hline
 780914& 60228.761&   0.125&  28.13&   139&    0.7&    1.1&    1.3&  0.72&   30.0&    8.4&  C \cr
 780918 & 71327.164&   0.125&  14.06&   142&    1.8&    1.9&    2.3&  1.13&   30.2&   11.4&  C  \cr
 780921 & 14222.567&   2.000&  14.06&   140&    1.4&    0.9&    1.4&  0.77&    4.4&    2.3&  CI  \cr
 781019 & 66357.776&        &  16.88&   138&    0.9&    1.4&    1.6&  1.00&    7.8&    3.3&  BDI  \cr
 781104 & 58627.977&   0.125&  28.08&   140&    3.7&    6.5&    7.1&  1.11&   18.8&    7.4&  C      \cr
 781115 & 76002.145&        &  11.25&   137&    1.1&    2.0&    1.9&  0.87&    2.5&    1.1&  BDI    \cr
 781119 & 34102.416&   8.000&  28.02&   139&    4.3&    7.5&    8.9&  1.12&   12.0&    3.6&  C      \cr
\end{longtable}
\pagebreak
\centerline{Table 1 (continued): Properties of PVO GRBs}
\begin{longtable}{l r r r r r r r r r r l}
\hline
\hline
Date &UT &$\Delta T_{\rm RT}$ &$T_{\rm MRO}$ &Bkg &1/4 s &1 s &4 s &TE &$T_{\rm 90}$ &$T_{\rm 50}$ &Notes \\
\hline
 800213 & 27052.207&        &  28.06&   110&    4.1&    6.3&   11.2&  1.11&   21.1&   13.1&  BDE    \cr
 800228 & 37993.727&   0.125&  19.69&   114&    0.4&    0.6&    0.9&  0.93&    1.0&    0.7&  CDH    \cr
 800322 & 26336.924&   2.000&  19.68&   110&    3.0&    3.2&    3.0&  1.10&    5.4&    3.7&  CI     \cr
 800331 &  2862.299&   2.000&  25.31&   110&    2.0&    3.2&    4.8&  1.10&   22.1&    5.4&  CI     \cr
 800408 & 51596.984&   2.000&   5.62&   103&    1.4&    0.6&    0.4&  0.38&    0.3&    0.0&  CI     \cr
 800410 & 75233.794&   2.000&  25.31&   106&    1.1&    1.4&    1.7&  0.59&   13.6&    5.1&  CI     \cr
 800412B& 72196.477&   2.000&  14.06&   108&    0.8&    1.7&    2.8&  0.13&  122.9&   20.7&  C      \cr
\end{longtable}

\pagebreak
\centerline{Table 1 (continued): Properties of PVO GRBs}
\begin{longtable}{l r r r r r r r r r r l}
\hline
\hline
Date &UT &$\Delta T_{\rm RT}$ &$T_{\rm MRO}$ &Bkg &1/4 s &1 s &4 s &TE &$T_{\rm 90}$ &$T_{\rm 50}$ &Notes \\
\hline
 811016 & 85923.086&        &  20.86&   509&    9.6&   12.6&   16.8&  1.00&   13.2&    4.5&  BD     \cr
 811209 & 11514.264&   2.000&   5.63&    96&    1.0&    1.5&    0.9&  0.81&   26.4&   11.6&  CI     \cr
 811215 & 71943.760&   2.000&  28.13&   100&    0.7&    1.1&    1.8&  0.76&   21.7&    8.2&  C      \cr
 811225 & 13524.775&   0.125&  22.50&    97&    0.9&    1.4&    1.8&  0.55&   16.8&    6.9&  CI     \cr
 811231B&  5789.168&        &  28.13&   177&    0.9&    1.3&    1.6&  1.01&    6.4&    3.5&  AD     \cr
 811231A& 28782.427&   2.000&   5.59&    94&    3.1&    1.5&    1.2&  1.12&    0.2&    0.1&  CI     \cr
 820104 & 57796.693&   0.125&   8.44&    94&    0.8&    1.2&    1.6&  1.13&    4.2&    1.6&  CDI    \cr
\end{longtable}

\pagebreak
\centerline{Table 1 (continued): Properties of PVO GRBs}
\begin{longtable}{l r r r r r r r r r r l}
\hline
\hline
Date &UT &$\Delta T_{\rm RT}$ &$T_{\rm MRO}$ &Bkg &1/4 s &1 s &4 s &TE &$T_{\rm 90}$ &$T_{\rm 50}$ &Notes \\
\hline
 831119 & 12801.339&   0.125&  22.50&   119&    1.6&    2.0&    2.1&  1.13&    6.3&    1.6&  CI     \cr
 831124 & 79061.955&        &  14.06&   116&    0.8&    1.3&    1.8&  0.56&    5.3&    2.2&  BDI    \cr
 831130 & 74319.281&   2.000&  28.13&   119&    1.1&    2.2&    3.3&  0.85&  154.4&   37.3&  C      \cr
 831221 & 54715.114&        &   5.61&   117&    2.0&    1.1&    0.6&  0.96&    0.2&    0.0&  BI     \cr
 840110 & 86128.850&        &  28.13&   117&    1.4&    2.5&    2.3&  0.83&   25.0&    5.3&  B      \cr
 840119 & 53254.695&   8.000&   5.55&   125&    3.2&    1.7&    1.0&  1.13&    0.8&    0.0&  CI     \cr
 840222 & 81893.843&        &  11.25&   128&    1.4&    1.6&    1.4&  1.10&    5.5&    1.1&  BDI    \cr
\end{longtable}
\pagebreak
\centerline{Table 1 (continued): Properties of PVO GRBs}
\begin{longtable}{l r r r r r r r r r r l}
\hline
\hline
Date &UT &$\Delta T_{\rm RT}$ &$T_{\rm MRO}$ &Bkg &1/4 s &1 s &4 s &TE &$T_{\rm 90}$ &$T_{\rm 50}$ &Notes \\
\hline
 850805 &  2924.284&   4.000&  28.03&   142&    5.1&    7.6&    9.9&  1.11&   21.2&    3.9&  C      \cr
 850811 &  1523.934&   0.250&  14.06&   140&    1.1&    1.2&    1.0&  0.72&   16.6&   13.5&  CI     \cr
 850820 & 53042.864&        &  14.06&   139&    2.4&    3.1&    3.6&  1.10&    9.4&    1.9&  BD     \cr
 850826 & 46472.431&   5.996&   8.44&   140&    1.0&    1.2&    1.5&  0.29&    3.9&    1.6&  CI     \cr
 850907 & 50712.095&   5.996&  14.06&   144&    0.9&    1.4&    1.3&  0.41&   10.0&    3.4&  C      \cr
 850922 & 45442.764&        &  28.13&   146&    0.9&    1.3&    2.1&  0.76&   16.7&    7.4&  BDE    \cr
 850927 & 42919.193&  12.000&  28.13&   147&    2.1&    3.7&    6.7&  1.03&   22.4&    9.2&  C      \cr
\end{longtable}
\pagebreak
\centerline{Table 1 (continued): Properties of PVO GRBs}
\begin{longtable}{l r r r r r r r r r r l}
\hline
\hline
Date &UT &$\Delta T_{\rm RT}$ &$T_{\rm MRO}$ &Bkg &1/4 s &1 s &4 s &TE &$T_{\rm 90}$ &$T_{\rm 50}$ &Notes \\
\hline
 880111B& 53905.994&   8.000&   2.81&   150&    0.5&    0.3&    0.6&  0.86&    4.0&    3.3&  CH     \cr
 880128 & 40559.515&   0.250&  28.04&   153&    3.7&    5.6&    6.9&  1.12&   12.7&    5.5&  C      \cr
 880205 & 37158.728&   2.000&  28.13&   155&    1.3&    2.2&    4.1&  0.69&   28.2&    9.9&  C      \cr
 880301 & 45495.146&        &  11.25&   152&    0.8&    1.2&    1.3&  0.59&    5.1&    1.7&  BDI    \cr
 880310 & 13069.198&   3.000&  28.13&   150&    2.1&    2.2&    2.5&  1.07&   21.8&    9.1&  C      \cr
 880311 & 79992.125&        &  23.94&   150&   13.4&   19.9&   23.9&  1.06&    9.0&    4.6&  B      \cr
 880318 & 59101.103&   2.000&  19.69&   152&    1.6&    2.7&    4.4&  0.71&   12.3&    4.4&  C      \cr
\end{longtable}

\pagebreak
\centerline{Table 1 (continued): Properties of PVO GRBs}
\begin{longtable}{l r r r r r r r r r r l}
\hline
\hline
Date &UT &$\Delta T_{\rm RT}$ &$T_{\rm MRO}$ &Bkg &1/4 s &1 s &4 s &TE &$T_{\rm 90}$ &$T_{\rm 50}$ &Notes \\
\hline
 890828 & 21500.393&   4.000&  27.69&   275&    4.7&    7.2&    8.2&  1.08&   14.5&    3.6&  C      \cr
 890915 &  5725.229&        &  28.13&    90&    2.4&    3.4&    4.6&  1.13&   14.4&    4.8&  B      \cr
 890923 & 30662.920&   8.000&  11.02&    95&    8.4&   12.2&    7.3&  1.00&    5.1&    0.8&  CI     \cr
 891012 & 40146.102&   8.000&  14.06&   122&    1.4&    2.5&    2.5&  0.87&    5.9&    1.7&  CI     \cr
 891207 & 64559.422&        &  14.06&    91&    1.2&    1.2&    1.2&  1.06&    6.5&    4.2&  CI     \cr
 891221 & 60542.517&        &  19.69&    76&    0.6&    0.9&    1.1&  0.90&    7.9&    3.8&  CHI    \cr
 900101 & 65452.923&        &  16.62&    74&    9.6&   10.4&    6.1&  1.06&    0.8&    0.5&  BI     \cr
\end{longtable}

\pagebreak
\centerline{Table 1 (continued): Properties of PVO GRBs}
\begin{longtable}{l r r r r r r r r r r l}
\hline
\hline
Date &UT &$\Delta T_{\rm RT}$ &$T_{\rm MRO}$ &Bkg &1/4 s &1 s &4 s &TE &$T_{\rm 90}$ &$T_{\rm 50}$ &Notes \\
\hline
 920212 & 15630.702&        &  25.52&    94&   12.6&   23.1&   36.4&  1.00&   14.0&    4.7&  B      \cr
 920226 & 29339.015&        &   8.44&    99&    0.9&    1.2&    1.2&  0.72&    2.8&    1.7&  BDG    \cr
 920311 &  8887.280&   0.375&  28.13&   102&    1.9&    2.9&    4.3&  1.12&   16.7&    7.0&  CG     \cr
 920325 & 61525.530&  16.000&  11.25&    99&    1.2&    1.0&    0.7&  0.81&   52.8&   24.9&  C      \cr
 920406 &  9756.428&        &  22.50&   105&    2.5&    4.1&    4.5&  1.11&   17.2&    4.0&  CG     \cr
 920414 & 84166.415&        &   8.44&   102&    0.9&    1.1&    0.7&  0.10&    3.0&    0.5&  BDG    \cr
 920419 &  4180.082&        &   5.63&   101&    1.0&    1.4&    1.0&  0.16&    1.6&    0.7&  BD     \cr
 920426 &  3956.730&        &  28.13&   103&    0.6&    0.9&    1.4&  0.19&   20.0&    8.8&  BDH    \cr
 920501 & 76840.660&        &  28.13&   107&    1.4&    1.3&    2.1&  1.13&   17.8&    6.9&  BD     \cr
 920517 & 12672.789&  16.000&  16.88&   100&    2.3&    2.2&    2.4&  1.13&    8.0&    1.3&  CG     \cr
 920519 & 59211.868&        &  28.13&   102&    0.9&    1.4&    1.4&  0.50&   20.7&   15.9&  BD     \cr
 920525 & 12881.961&  16.000&  22.50&   110&    2.3&    3.6&    2.4&  1.13&   13.7&    9.2&  CG     \cr
 920711 & 58076.640&        &  28.13&   108&    1.2&    2.1&    2.5&  0.66&   23.5&    9.0&  BD     \cr
 920720 & 11155.341&   8.000&  11.25&   112&    1.2&    1.3&    1.2&  1.12&    7.7&    4.5&  CG     \cr
 920723 & 72694.102&        &  19.55&   109&    6.0&    9.6&    9.8&  1.10&    4.8&    2.5&  B      \cr
 920731 & 61857.429&        &  28.13&   110&    1.0&    1.4&    1.7&  0.94&   24.3&   18.3&  BD     \cr
 921004 & 49963.747&        &  28.11&   107&    3.8&    6.4&    5.6&  1.10&   12.1&    8.8&  B      \cr
\hline
\end{longtable}

\noindent Footnotes: \hfill\break
A: Portions of MRO used to determine background
\hfill\break
B: Background from non-simulataneous RT
\hfill\break
C: Background from near simulataneous RT
\hfill\break
D: Data availibility makes $T_{90}$ and $S$ unreliable
\hfill\break
E: Data availibility makes $P$ and $C_{\rm max}$ unreliable
\hfill\break
F: $\sigma_t$ set to 5.6
\hfill\break
G: Also seen by BATSE
\hfill\break
H: Should not be used in Log~$N$-Log~$P$
\hfill\break
I: MRO terminated early

\pagebreak
\centerline{Table 2: PVO GRBs Peak Luminosity and Fluence}
\begin{longtable}{l r r r r r r r r r }
\hline
\hline
Date &$P_{\rm P}$ &$S_{\rm P}$ &$P_{\rm B}$ &$S_{\rm B}$ &$P_{\rm T}$ &$S_{\rm T}$ &HR$_{\rm P}$ &$\sigma_{\rm HR}$ &HR$_{\rm S}$  \\
\hline
  780914  &    25.1 &    3.57 &   23.0 &    24.5 &    3.02 &    2.74 &     1.5 &    0.68 &     2.0 \cr
  780918  &    55.5 &    8.62 &   36.6 &    51.5 &    3.70 &    3.46 &     1.9 &    0.44 &     4.8 \cr
  780921  &    44.0 &    3.45 &   14.7 &    34.8 &    0.91 &    1.18 &     4.1 &    1.68 &     8.6 \cr
  781019  &    31.8 &    5.70 &   36.0 &    36.8 &    1.09 &    1.42 &     1.4 &    0.49 &    11.8 \cr
  781104  &   112.8 &   36.61 &   48.7 &    89.2 &    6.73 &    9.30 &     3.3 &    0.60 &    13.2 \cr
  781115  &    41.8 &    5.22 &   24.4 &    33.9 &    0.95 &    1.28 &     2.9 &    1.01 &    12.7 \cr
  781119  &   117.6 &   39.16 &   32.1 &    93.0 &    4.17 &    8.66 &     3.9 &    0.69 &    31.5 \cr
\end{longtable}
\pagebreak
\centerline{Table 2 (continued): PVO GRBs Peak Luminosity and Fluence}
\begin{longtable}{l r r r r r r r r r }
\hline
\hline
Date &$P_{\rm P}$ &$S_{\rm P}$ &$P_{\rm B}$ &$S_{\rm B}$ &$P_{\rm T}$ &$S_{\rm T}$ &HR$_{\rm P}$ &$\sigma_{\rm HR}$ &HR$_{\rm S}$  \\
\hline
  800213  &   116.3 &   72.44 &   63.4 &    94.5 &   18.42 &   21.61 &     2.4 &    0.38 &     8.6 \cr
  800228  &    10.8 &    0.87 &   32.0 &    32.4 &    0.11 &    0.19 &     0.5 &    0.30 &    20.8 \cr
  800322  &    82.1 &    7.14 &   72.9 &    88.6 &    1.78 &    2.30 &     1.7 &    0.29 &     9.0 \cr
  800331  &    59.8 &   13.50 &   94.8 &    83.8 &   11.05 &    9.35 &     0.9 &    0.18 &     2.6 \cr
  800408  &    35.6 &    0.82 &   11.3 &    28.1 &    0.12 &    0.20 &     3.1 &    1.25 &    19.4 \cr
  800410  &    27.8 &    4.92 &   17.0 &    26.3 &    2.12 &    1.65 &     5.3 &    3.01 &     4.7 \cr
  800412B &    24.4 &   57.36 &   10.7 &    19.3 &    7.82 &   12.68 &     5.1 &    3.75 &    19.0 \cr
\end{longtable}
\pagebreak
\centerline{Table 2 (continued): PVO GRBs Peak Luminosity and Fluence}
\begin{longtable}{l r r r r r r r r r }
\hline
\hline
Date &$P_{\rm P}$ &$S_{\rm P}$ &$P_{\rm B}$ &$S_{\rm B}$ &$P_{\rm T}$ &$S_{\rm T}$ &HR$_{\rm P}$ &$\sigma_{\rm HR}$ &HR$_{\rm S}$  \\
\hline
  811016  &   558.7 &  128.80 &  418.8 &   522.3 &   49.38 &   52.34 &     1.8 &    0.11 &     5.1 \cr
  811209  &    24.3 &    2.25 &   24.0 &    27.0 &    1.68 &    1.58 &     2.2 &    0.88 &     2.5 \cr
  811215  &    22.9 &    5.77 &   22.3 &    26.9 &    3.84 &    3.80 &     1.9 &    0.82 &     2.8 \cr
  811225  &    23.3 &    4.04 &   18.5 &    24.2 &    2.08 &    1.87 &     2.0 &    0.82 &     3.9 \cr
  811231B &    27.0 &    7.14 &   40.0 &    33.9 &    1.67 &    2.08 &     0.9 &    0.34 &     9.8 \cr
  811231A &   101.9 &    2.35 &   25.6 &    80.6 &    0.20 &    0.52 &     2.9 &    0.55 &    42.3 \cr
  820104  &    25.9 &    2.25 &   52.6 &    42.0 &    1.07 &    0.92 &     0.7 &    0.27 &     4.0 \cr
\end{longtable}
\pagebreak
\centerline{Table 2 (continued): PVO GRBs Peak Luminosity and Fluence}
\begin{longtable}{l r r r r r r r r r }
\hline
\hline
Date &$P_{\rm P}$ &$S_{\rm P}$ &$P_{\rm B}$ &$S_{\rm B}$ &$P_{\rm T}$ &$S_{\rm T}$ &HR$_{\rm P}$ &$\sigma_{\rm HR}$ &HR$_{\rm S}$  \\
\hline
  831119  &    41.8 &    4.62 &   41.0 &    47.9 &    2.12 &    2.06 &     2.4 &    0.72 &     4.3 \cr
  831124  &    23.0 &    4.62 &   18.2 &    24.6 &    1.30 &    1.53 &     4.5 &    2.99 &     7.4 \cr
  831130  &    33.3 &   49.21 &   23.5 &    32.7 &   15.07 &   17.40 &     2.3 &    0.88 &     6.7 \cr
  831221  &    56.3 &    1.39 &   23.1 &    44.5 &    0.19 &    0.33 &     4.1 &    1.20 &    21.5 \cr
  840110  &    39.6 &   12.63 &   22.9 &    31.3 &    1.25 &    2.79 &     2.4 &    0.71 &    34.5 \cr
  840119  &    90.2 &    2.29 &   29.6 &    71.3 &    0.27 &    0.51 &     3.6 &    0.88 &    27.4 \cr
  840222  &    43.0 &    3.22 &   36.7 &    47.3 &    0.88 &    1.07 &     2.0 &    0.55 &     7.9 \cr
\end{longtable}
\pagebreak
\centerline{Table 2 (continued): PVO GRBs Peak Luminosity and Fluence}
\begin{longtable}{l r r r r r r r r r }
\hline
\hline
Date &$P_{\rm P}$ &$S_{\rm P}$ &$P_{\rm B}$ &$S_{\rm B}$ &$P_{\rm T}$ &$S_{\rm T}$ &HR$_{\rm P}$ &$\sigma_{\rm HR}$ &HR$_{\rm S}$  \\
\hline
  850805  &   136.6 &   43.04 &   29.8 &   108.0 &    5.23 &    9.51 &     4.6 &    0.77 &    25.1 \cr
  850811  &    34.8 &    6.98 &   19.5 &    28.1 &    1.79 &    2.19 &     1.9 &    0.67 &     8.3 \cr
  850820  &    71.2 &   10.27 &   27.0 &    56.3 &    2.18 &    2.68 &     3.9 &    1.02 &    10.3 \cr
  850826  &    32.2 &    3.52 &   10.6 &    25.5 &    0.72 &    0.96 &     2.5 &    0.95 &    11.0 \cr
  850907  &    28.0 &    5.81 &   18.1 &    22.1 &    0.78 &    1.29 &     2.0 &    0.84 &    19.4 \cr
  850922  &    27.4 &   12.92 &   22.9 &    26.3 &    4.10 &    4.57 &     1.9 &    0.79 &     6.4 \cr
  850927  &    69.5 &   73.30 &   21.4 &    54.9 &    9.64 &   16.20 &     4.2 &    1.28 &    22.3 \cr
\end{longtable}
\pagebreak
\centerline{Table 2 (continued): PVO GRBs Peak Luminosity and Fluence}
\begin{longtable}{l r r r r r r r r r }
\hline
\hline
Date &$P_{\rm P}$ &$S_{\rm P}$ &$P_{\rm B}$ &$S_{\rm B}$ &$P_{\rm T}$ &$S_{\rm T}$ &HR$_{\rm P}$ &$\sigma_{\rm HR}$ &HR$_{\rm S}$  \\
\hline
  880111B &    29.9 &    2.69 &   24.4 &    32.9 &    0.22 &    0.59 &     3.2 &    1.36 &    41.3 \cr
  880128  &   113.8 &   45.14 &   35.0 &    90.0 &    5.21 &    9.98 &     4.9 &    1.05 &    28.9 \cr
  880205  &    44.5 &   45.46 &   11.4 &    35.2 &    3.35 &   10.05 &     3.2 &    1.12 &    58.4 \cr
  880301  &    28.9 &    4.44 &   20.5 &    22.8 &    0.69 &    0.98 &     2.2 &    0.94 &    15.8 \cr
  880310  &    67.0 &   21.73 &   26.8 &    53.0 &    3.70 &    5.06 &     3.5 &    1.00 &    14.0 \cr
  880311  &   427.3 &  119.20 &  220.9 &   353.5 &   20.90 &   28.57 &     3.1 &    0.23 &    14.9 \cr
  880318  &    47.6 &   24.65 &    9.7 &    37.6 &    2.05 &    5.45 &     6.5 &    3.06 &    51.2 \cr
\end{longtable}
\pagebreak
\centerline{Table 2 (continued): PVO GRBs Peak Luminosity and Fluence}
\begin{longtable}{l r r r r r r r r r }
\hline
\hline
Date &$P_{\rm P}$ &$S_{\rm P}$ &$P_{\rm B}$ &$S_{\rm B}$ &$P_{\rm T}$ &$S_{\rm T}$ &HR$_{\rm P}$ &$\sigma_{\rm HR}$ &HR$_{\rm S}$  \\
\hline
  890828  &   205.2 &   51.13 &  103.7 &   162.2 &    9.05 &   12.23 &     2.5 &    0.30 &    13.7 \cr
  890915  &    65.5 &   17.68 &   39.3 &    55.4 &    4.13 &    5.13 &     2.3 &    0.53 &     9.4 \cr
  890923  &   224.1 &    9.33 &  481.9 &   389.6 &   11.08 &   10.14 &     0.7 &    0.07 &     1.4 \cr
  891012  &    43.3 &    8.54 &   23.9 &    34.2 &    1.61 &    2.05 &     2.4 &    0.73 &    12.1 \cr
  891207  &    29.2 &    1.58 &   39.9 &    39.8 &    2.19 &    1.98 &     1.3 &    0.42 &     1.3 \cr
  891221  &    16.6 &    2.39 &   33.6 &    27.8 &    0.91 &    1.02 &     0.7 &    0.36 &     5.4 \cr
  900101  &   232.6 &    6.12 &  255.0 &   273.6 &    3.38 &    3.44 &     1.5 &    0.14 &     3.4 \cr
\end{longtable}
\pagebreak
\centerline{Table 2 (continued): PVO GRBs Peak Luminosity and Fluence}
\begin{longtable}{l r r r r r r r r r }
\hline
\hline
Date &$P_{\rm P}$ &$S_{\rm P}$ &$P_{\rm B}$ &$S_{\rm B}$ &$P_{\rm T}$ &$S_{\rm T}$ &HR$_{\rm P}$ &$\sigma_{\rm HR}$ &HR$_{\rm S}$  \\
\hline
  920212  &   323.0 &   69.03 &  640.3 &   531.0 &   71.59 &   63.69 &     0.7 &    0.06 &     1.8 \cr
  920226  &    23.9 &    2.13 &   21.3 &    26.2 &    0.59 &    0.73 &     2.2 &    0.99 &     8.2 \cr
  920311  &    53.2 &   18.97 &   49.0 &    58.1 &    5.08 &    6.10 &     1.7 &    0.40 &     8.0 \cr
  920325  &    39.3 &    5.26 &   18.5 &    32.5 &    1.24 &    1.54 &     2.8 &    0.97 &     9.0 \cr
  920406  &    72.3 &   20.63 &   66.7 &    79.3 &    5.56 &    6.56 &     1.7 &    0.33 &     8.0 \cr
  920414  &    22.0 &    1.42 &   10.4 &    18.2 &    0.40 &    0.36 &     1.9 &    0.87 &     7.1 \cr
  920419  &    26.8 &    2.18 &   10.1 &    21.2 &    0.21 &    0.48 &     3.1 &    1.45 &    36.5 \cr
  920426  &    15.6 &    5.59 &   20.3 &    12.4 &    2.27 &    2.41 &     1.5 &    0.81 &     4.9 \cr
  920501  &    38.6 &    8.65 &   48.6 &    48.5 &    4.11 &    4.28 &     1.2 &    0.30 &     4.1 \cr
  920517  &    61.5 &    5.63 &   43.4 &    54.8 &    1.90 &    2.06 &     2.4 &    0.54 &     6.1 \cr
  920519  &    27.9 &   10.38 &   17.1 &    24.7 &    2.14 &    2.89 &     2.5 &    1.00 &    11.4 \cr
  920525  &    61.5 &   13.79 &   44.2 &    50.6 &    3.31 &    3.95 &     2.0 &    0.40 &     9.0 \cr
  920711  &    33.5 &   17.80 &   18.6 &    26.9 &    3.11 &    4.48 &     3.4 &    1.43 &    14.4 \cr
  920720  &    37.9 &    3.55 &   40.3 &    45.8 &    1.38 &    1.41 &     1.9 &    0.65 &     5.1 \cr
  920723  &   158.1 &   33.59 &   64.1 &   125.0 &    4.69 &    7.43 &     3.6 &    0.50 &    20.2 \cr
  920731  &    34.4 &   12.48 &   29.1 &    36.9 &    4.09 &    4.74 &     1.9 &    0.60 &     6.4 \cr
  921004  &   108.4 &   18.26 &   90.3 &   110.1 &    5.07 &    6.03 &     2.0 &    0.30 &     7.7 \cr
\hline
\end{longtable}

\noindent
References:

Band, D. L., et al. 1993, ApJ, 413, 281.

Costa, E., et al.  1997, Nature, 387, 783.

Fenimore, E. E., er al. 1993, Nature. 366, 40.

Klebesadel, R. W., Strong, I. B., and Olson, R. A., 1973, Ap. J. Letters 182, L85.

Klebesadel, R., Evans, W., Glore, J., Spalding, R., and Wymer, F., 1980 IEEE Trans on Geoscience and Remote Sensing, GE-18, 76.

\end{document}